# A Noninvasive Magnetic Stimulator Utilizing Secondary Ferrite Cores and Resonant Structures for Field Enhancement

Raunaq Pradhan[1] and Yuanjin Zheng [1]

[1]School of Electrical and Electronic Engineering, Nanyang Technological University, Singapore

**Abstract:** *In this paper, secondary ferrite cores and resonant structures have been used for field enhancement. The tissue was placed between the double square source coil and the secondary ferrite core. Resonant coils were added which aided in modulating the electric field in the tissue. The field distribution in the tissue was measured using electromagnetic simulations and ex-vivo measurements with tissue. Calculations involve the use of finite element analysis (Ansoft HFSS) to represent the electrical properties of the physical structure. The setup was compared to a conventional design in which the secondary ferrite cores were absent. It was found that the induced electric field could be increased by 122%, when ferrite cores were placed below the tissue at 450 kHz source frequency. The induced electric field was found to be localized in the tissue, verified using ex-vivo experiments. This preliminary study maybe further extended to establish the verified proposed concept with different complicated body parts modelled using the software and in-vivo experiments as required to obtain the desired induced field.*

Corresponding author: Raunaq Pradhan; email: raunaq.pradhan@gmail.com; phone: +65-83132872

## I.    INTRODUCTION

The development of ferrite cores which could be used for power applications at kHz frequencies have helped in the manufacture of magnetic stimulators working at these frequencies. Ferrite cores have been utilized to increase the electric field in the nerve. Magnetic stimulators provide a spatial rate of change of electric field in the tissue by producing a magnetic field created by a varying current flowing through the primary coil. The major advantages of this type of stimulation include minimal discomfort to patients and its noncontact, noninvasive nature. Clinical applications predominantly use the Figure of Eight (FOE) coils. Other topologies like the double square coils, Cadwell coils, and the quad square coils also exist. Literature estimates that larger activation per unit current is produced when quad and square coils are used [1]. For low frequency magnetic stimulation, air cores have been used.

Power consumption for these magnetic stimulators, which consists of Transcranial Magnetic Stimulation (TMS) and repetitive Transcranial Magnetic Stimulation (rTMS) modalities are very high. The study of pulsed electromagnetic fields in human tissues required the scaling down of the stimulation systems. One such work involves the use of ferrite cores in quad coils at frequencies between 200 kHz - 1 MHz to achieve stimulation [2]. The stimulation depends on three parameters, 1) the orientation of the coil with respect to the head or body as the case may be, 2) the current waveform through the coil and 3) the type of coil.

Since Baker introduced the use of magnetic fields to stimulate the human motor cortex and the peripheral nerves, which can produce effects like muscle twitches; there has been tremendous interest in the medical field. Design of full scale magnetic stimulation systems consists of many challenges. During the discharge phase, a magnetic stimulator typically produces an output power of 5 MW. In the case of TMS/rTMS, this would imply that a capacitor needs to convert the charge stored into a magnetic energy in 100µs [3]. When higher frequencies are chosen, this results in stringent requirements for capacitor selection and eventually becomes a bottle-neck. Extensive work on the use of rTMS and TMS for curing various diseases like visceral pain, chronic neuropathic pain and fibromyalgia has been presented in literature. These modalities involve passing high intensity current pulses through a coil of wire, thereby inducing a magnetic field as high as 2-4 T [4]. Fast changing currents can be produced by two approaches, namely, using the SCR (silicon controlled rectifier) based charging and discharging circuit, controlled by a PWM input, using similar circuits as IGBTs [5-6].

Increasing the field localization for stimulation has been yet another field of research. 3D finite element simulation has been used to determine the field localization and the electric field at the point of stimulation. Stimulation is carried out by using a FOE coil, with a conductive plate having a window placed under it. The field localization was improved due to the addition of the conductive plate, but the electric field in the tissue decreased 2 fold for the same FOE coil dimensions [7].

This paper proposes a setup which utilizes the use of ferrite cores and resonant structures to enhance the induced electric field, and compares it to a conventional design. A tissue model is used to simulate the stimulation, followed by ex-vivo experiments using pork tissue, where the findings have been validated.

## II.    BACKGROUND THEORY

### A) Frequency selection for stimulation

Recently there have been studies which estimate the effects of magnetic fields and time varying currents from kHz to MHz frequencies. It has also been found that the membrane voltage increased with increase in frequency of the applied magnetic field [8]. The increase in membrane voltage is considered enough evidence to perform frequency selection for magnetic stimulation. It should be noted that at lower frequencies, the induced electric field in the tissue is smaller. The membrane voltage alone does not help in determining the suitable value for stimulation. It is shown in literature that the nerve acts as a low pass filter and internally amplifies low frequency signals compared to high frequency signals. This is because the nerve consists of capacitors and resistors which help in functioning as a low pass filter with a large DC gain. At higher frequencies capacitors tend to act as short circuits lowering the DC gain. The initial estimate of the electric field's amplification factor is given by the following equation:

$$G_E(\omega) = \frac{E_m(\omega)}{E_e} = \frac{3R_c}{2d} \frac{1}{1 + j\omega\tau_m} \tag{1}$$

*Where*

$$\tau_m = \frac{R_c C_m}{\frac{2\lambda_i\lambda_e}{\lambda_i + \lambda_e} + \frac{R}{d}\lambda_m} \tag{2}$$

Where, $R_c$ is the cell radius, $d$ is the membrane thickness which is equal to $3\times10^{-9}$, $\lambda_i$ is the conductivity of the cytoplasm, $\lambda_m$ is the conductivity of the membrane, $\lambda_e$ is the conductivity of the extracellular. $E_m$ is the electric field inside the membrane. However, this model does not consider the second order effects. $\varepsilon_i, \varepsilon_m, \varepsilon_e$ are the permittivity of the cytoplasm, the membrane and the extracellular medium respectively. $\varepsilon_e$ was taken as 77.8 and $\varepsilon_m$ was taken as 3.95. A more precise version of the amplification factor is given below.

$$G_E(\omega) = \frac{3\lambda_e^*\left[3R_c^3\lambda_i^* + (3dR_c^2 - d^2R_c)(\lambda_m^* - \lambda_i^*)\right]}{K - 2(R_c - d)^3(\lambda_e^* - \lambda_m^*)(\lambda_i^* - \lambda_m^*)} \tag{3}$$

Where,

$$K = 2R_c^3(\lambda_m^* - 2\lambda_e^*)\left(\lambda_m^* - \frac{1}{2}\lambda_e^*\right). \tag{4}$$

where, $\lambda^* = \lambda + j\omega\varepsilon$. Here, second order effects are also considered. The nerve amplification function generates a zero at 100 MHz, the effects felt by the zero were not considered in the first order solution. The above calculation is not exact because it considers the nerve as a cylinder [9-11].

Exact formulation is tedious because the exact geometry of the nerve is to be incorporated and solved by numerical methods and hence is not considered in this work. By substituting the extracellular medium, the amplification factor was plotted to get an approximate estimate of the frequency range to be chosen. Four extracellular mediums, namely blood, fat, bone and muscle where chosen for the estimation for which the conductivities are obtained from literature [11-12]. The logarithmic magnitude of the amplification factor and the log magnitude of the frequency were plotted to obtain Fig 1. It was seen that the decrease in the amplification factor was predominant for frequencies greater than 1 MHz In the case of fat tissue, the amplification factor had significant reduction even before 1 MHz; hence frequencies in the order of hundreds of kHz were chosen for analysis.

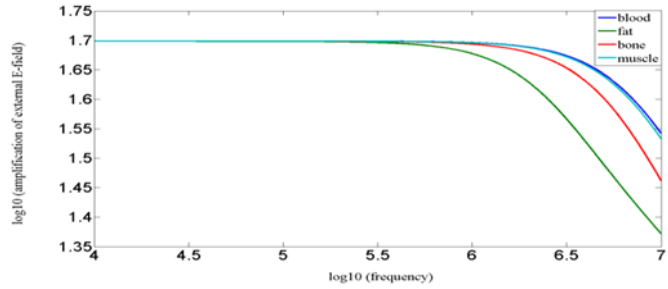

Fig. 1. Amplification factor as a function of frequency

## B) Primary Electric field due to the coils

Neural stimulation can be achieved by a spatial change of electric field along the nerve. It can be expressed as

$$\lambda_m{}^2 \frac{\partial \overrightarrow{E_x}(x,t)}{\partial x} = \tau \frac{\partial V_m(x,t)}{\partial t} - \lambda_m{}^2 \frac{\partial^2 V_m(x,t)}{\partial x^2} + V_m(x,t). \quad (5)$$

where the length and time constants are defined as

$$\lambda_m = \sqrt{\frac{r_m}{r_i}} \text{ and } \tau_m = c_m r_m. \qquad (6)$$

Where, x is the distance along the axis of the nerve fibre when the nerve fibre is aligned along the x axis. $E_x$ is the electric field along x axis. $V_m$ is the trans-membrane voltage which is the voltage difference between the intracellular and extracellular fluid. $r_m$ and $r_i$ is the membrane resistance times unit length and the intracellular resistance respectively. $c_m$ is the membrane capacitance per unit length. $\partial E_x / \partial x$ is the activation function.

The primary electric field generated due to the magnetic field created by a rate of change of current through the coil is given by the following equation:

$$\overrightarrow{dE^p} = \frac{-\mu_0 N \left(\frac{di}{dt}\right) \overrightarrow{dl}}{4\pi R}. \qquad (7)$$

where, $dE^p$ is the primary electric filed, $\mu_0$ is the permeability of free space, $\frac{di}{dt}$ is the rate of change of electric current, $dl$ is an element of the coil, $N$ is the number of turns and $R$ is the distance between the coil element and the point where the electric field is calculated.

For generating the required electric field in the nerve, various types of coils have been used for magnetic stimulation in literature. This is because magnetic stimulation provides the following advantages: 1) no contact to skin and non-invasive, 2) Ability to move and control the point of stimulation with ease. Though other coil structures have also been used, larger activating function per unit area has been achieved when square and quad coils are used. Electric field induced in the tissue can be calculated from the equation

$$\vec{E}_{Source}(\vec{r}) = -\frac{\partial I_1}{\partial t}\frac{\mu_0}{4\pi} \int_S \frac{\overrightarrow{dl'}_{Source}}{|\vec{r} - \vec{r'}|} - \nabla\emptyset_1. \qquad (8)$$

where, $\vec{E}_{Source}$ is the electric field produced in the tissue due to a current carrying source coil, $dl'$ is an infinitesimal coil section and $r'$ is the vector from each section to the point $r$. $\emptyset_1$ is the electric potential due to surface charge accumulation. For our calculations here, a double square coil is used. Adopting the methods proposed [1], we obtain the following relations for activation function.

$$\frac{\partial E_x}{\partial x} = -\frac{\mu_{eq}}{4\pi}\left\{N\left(\frac{dI_1}{dt}\right)(-A+B+C-D-E+F)\right\} - \nabla\emptyset_1 \qquad (9)$$

where

$$A = \frac{2}{\sqrt{x^2 + y^2 + (h-p)^2}}$$

$$B = \frac{1}{\sqrt{(s-x)^2 + y^2 + (h-p)^2}}$$

$$C = \frac{1}{\sqrt{(s+x)^2 + y^2 + (h-p)^2}}$$

$$D = \frac{1}{\sqrt{(s-x)^2 + (s-y)^2 + (h-p)^2}}$$

$$E = \frac{1}{\sqrt{(s+x)^2 + (s+y)^2 + (h-p)^2}}$$

$$F = \frac{2}{\sqrt{x^2 + (s-y)^2 + (h-p)^2}} \qquad (10)$$

where, $x$ and $y$ are the co-ordinates in the plane of the coil. $s$ is the length of a side of the square coil. $h$ is the total distance between the source and the receiver coil. $p$ is the point at which the electric field is measured. $N$ denotes the number of turns in the primary coil.

## C) Use of Resonant Structures for field enhancement

Now, it is to be estimated if a secondary structure is able to help the increase of the electric field at the nerve. We know from Maxwell's equations that

$$\oint E.\,dl = -\frac{\partial}{\partial t}\int_s B.\,ds. \qquad (11)$$

Where, $E$ is the electric field and $B$ is the magnetic field. Ferrites can be used as flux concentrators, thereby increasing the flux in the path between the primary and the secondary ferrite. This leads to the increase in electric field induced at the tissue.

Soft ferrites are hard, brittle and chemically inert. These materials are black or dark-grey. Compounds of either Manganese/Zinc or of Nickel/Zinc are primarily used. Magnetic properties are exhibited if below the Curie temperature. They possess high resistivity and high permeability. Composition of the oxides can be adjusted to tune the permeability during manufacturing process. Addition of resonating structures help in modifying the electric field. Rate of change of current in the primary coil can cause currents to flow in the secondary coil. This provides a change in the field distribution, thereby modifying the electric field to a small extent.

Wireless power transfer experiments were first carried out by Nikola Tesla. Recent interest in wireless power transfer is due to the work reported which used the method of resonant coupling for mid-range wireless power transfer. This was based on the understanding that a physical system can be reduced to a set of differential equations by using coupled mode theory [13].

$$\dot{a}_S = -i(\omega_S - i\Gamma_S)a_S - ika_D + Fe^{-i\omega t} \qquad (12)$$

$$\dot{a}_D = -i(\omega_D - i\Gamma_D)a_D - ika_S \qquad (13)$$

where, $\omega_S$ and $\omega_D$ are the resonant frequencies of the isolated objects and $\Gamma_S$ and $\Gamma_D$ are the intrinsic decay rates due to absorption and radiated losses. $k$ is the coupling co-efficient. $F$ is the driving term and $a$ is a variable defined so that the energy contained in the object is $|a_{S,D}(t)|^2$. When such a resonating structure is placed, it will generate an electric field which is given by:

$$\vec{E}_{receiver}(\vec{r}) = -\frac{\partial I_2}{\partial t}\frac{\mu_0}{4\pi}\int_S \frac{d\vec{l'}_{receiver}}{|\vec{r} - \vec{r''}|} - \nabla\emptyset_2 \qquad (14)$$

Assuming that a current flows in the secondary structure due to resonance, the electric field created by it on the nerve is given by:

$$\frac{\partial E_x}{\partial x} = -\frac{\mu_{eq}}{4\pi}\left[M\left(\frac{dI_2}{dt}\right)\left(\frac{1}{\sqrt{(x_0 - x)^2 + (y_0 - y)^2 + p^2}}\right)\right] \qquad (15)$$

This field has a vectorial additive effect to the electric field produced due to the primary coil, thereby modulating the electric field in the nerve. Here, $x_0$ and $y_0$ refer to points on the receiver coil. $M$ is used to refers to the number of turns in the receiver coil. $\mu_{eq}$ is the effective permittivity [2].

## III.  Materials and Methods

### A) Simulation Set-up

The Ansoft Simulation setup, shown below in Fig. 2 consisted of a primary coil (double square coil) for focusing the electric field to the tissue. The primary coil had ferrite cores for concentrating the flux. The secondary structure was varied as ferrite cores, ferrite cores and resonating/non-resonating coil, double square coils and vacuum. A pork fat tissue was modelled using the software (where conductivity values were taken from literature mentioned below [12]) was placed in between the primary coil and the secondary structure. The structure was enclosed in an air box. Finite Element-Boundary Integral (FE-BI) method was used to solve the structure. The simulation setup used in Ansoft HFSS [14] is shown in Fig. 2.

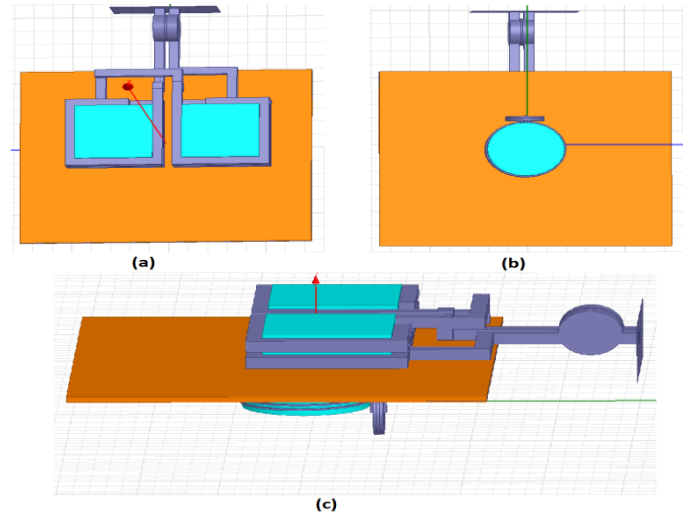

Fig. 2.   a) Top view of the simulation setup, b) Bottom view of the simulation setup, c) Side view of the simulation setup. Ferrites, fat tissue, copper are shown in light blue, brown and grey respectively.

FE-BI method available in HFSS was used because of the wavelength of the fields being in the order of hundreds of kilometers and hence using the ABC boundary condition which requires setting the air box length as $\lambda/4$ would be futile, due to large time for processing. Air box size was chosen as 11.1x11.1x11.1 m3. The values of $\varepsilon_r$, $\mu_r$ and $\sigma$ of the fat tissue were 31.565, 1 and 0.024906 S/m respectively. The values of $\varepsilon_r$, $\mu_r$ and $\sigma$ for the ferrite were 12, 1000 and 0.01 Siemens/m respectively and the values of $\varepsilon_r$, $\mu_r$ and $\sigma$ of copper were taken as 1, 0.999991 and 5.8x107 Siemens/m. The fat tissue, the air box, the secondary coil, the ferrite cores, the input capacitor, the secondary ferrite core, secondary capacitor and the primary coil had 1912, 45996, 3577, 935, 6708, 912, 117 and 4066 tetrahedron mesh elements respectively at 550 kHz. An input power of 500 W was provided to the source coil. The fat tissue had a thickness of 0.25 cm. Fields 0.2 cm below the tissue were analyzed.

## B) Parameter estimation based on circuit calculations

The parameters were chosen based on theoretical circuit model calculations and used as parameters for software based electromagnetic computation. The calculations of the parameter values are shown below.

1) Receiver coil resistance (R)$_r$: Because of skin effect, the current density on the surface of the wire is higher than that at the center of the wire. This is taken into consideration while calculating the resistance of the wire. The resistance is given as follows:

$$R_r = \frac{2\pi a M}{\sigma 2\pi r \delta} \qquad (16)$$

Where, a is the radius of the coil, M is the number of turns, r is the cross sectional area of the wire (0.071 cm), σ is the conductivity of the wire. The material chosen for the wire is copper and so σ=5.8·10$^7$. δ is the skin depth [15].

$$\delta = \frac{1}{\sqrt{\pi f \mu_0 \sigma}} \qquad (17)$$

where, $\mu_0$ is the permeability of free space ($\mu_0$=4π·10$^{-7}$). Solving for δ and substituting in (16), we obtain the resistance values at 450 kHz, 500 kHz and 550 kHz for the wires of radius 2 cm and 3 cm.

2) Coil resistance $R_L$: The measurement. is done by taking $R_L$ as 50 Ω. $R_r$ is negligible compared to $R_L$.

3) Source coil resistance $R_s$: The source coil is a double-square coil with each side, s=5 cm. Number of turns, N=1. Number of coils is 2, because the coil is a double-square coil. D is the width of the wire.

$$R_s = Number\ of\ coils \cdot \frac{4sN}{\sigma 4D\delta} \qquad (18)$$

which gives $R_s$=1.75 mΩ at 450 kHz, 1.845 mΩ at 500 kHz and 1.935 mΩ at 550 kHz. It should be noted that the calculations in this section do not take into consideration the effects caused due to the ferrites. Calculations of resistances for the copper coils are carried out.

## C) Ex-vivo experimental set-up

Besides, validation of the simulations, experiments for determining the electric field induced due to the stimulation set-up has also been carried out using a pork tissue with the dimensions of 2 cm x 2.5 cm x 2 cm. Here, the voltage across the tissue was measured using electrodes, both with and without the presence of ferrite cores. The primary coil is used as double square coils, whereas for the secondary structures, resonant structures have been used. The secondary ferrite was chosen as a cylinder. The motivation behind using this simple experimental set-up is to test the designed coils for their efficacy in achieving the required electric field enhancement via the use of proposed resonant coil structures and ferrite core structures at the desired site of stimulation. The voltage induced across the pork tissue connected with metal electrodes is observed using an oscilloscope. The set-up is described in Fig. 3.

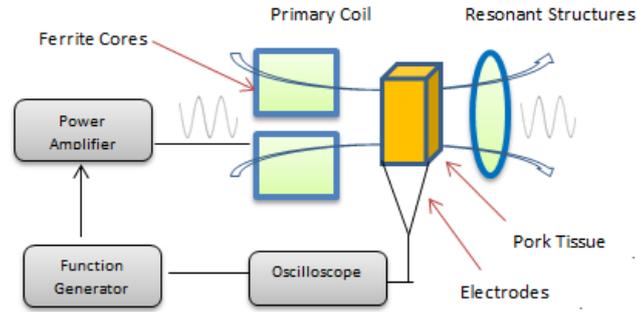

Fig. 3. Ex-vivo experimental set-up to measure voltage across electrodes for the simulated set-up

## IV.    Results and Discussion

Fig. 4. shows the maximum electric field observed in the tissue when the radius of the ferrite core is varied. In Fig. 4, for calculation it should be noted that the secondary structure is only a ferrite core. It was observed that when the radius of the ferrite core was chosen as 3 cm, the maximum magnitude of induced electric field in the tissue was obtained.

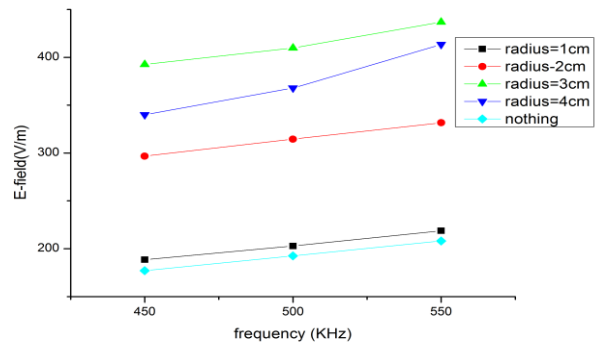

Fig. 4. Variation of maximum electric field with respect to

frequency for different ferrite radius

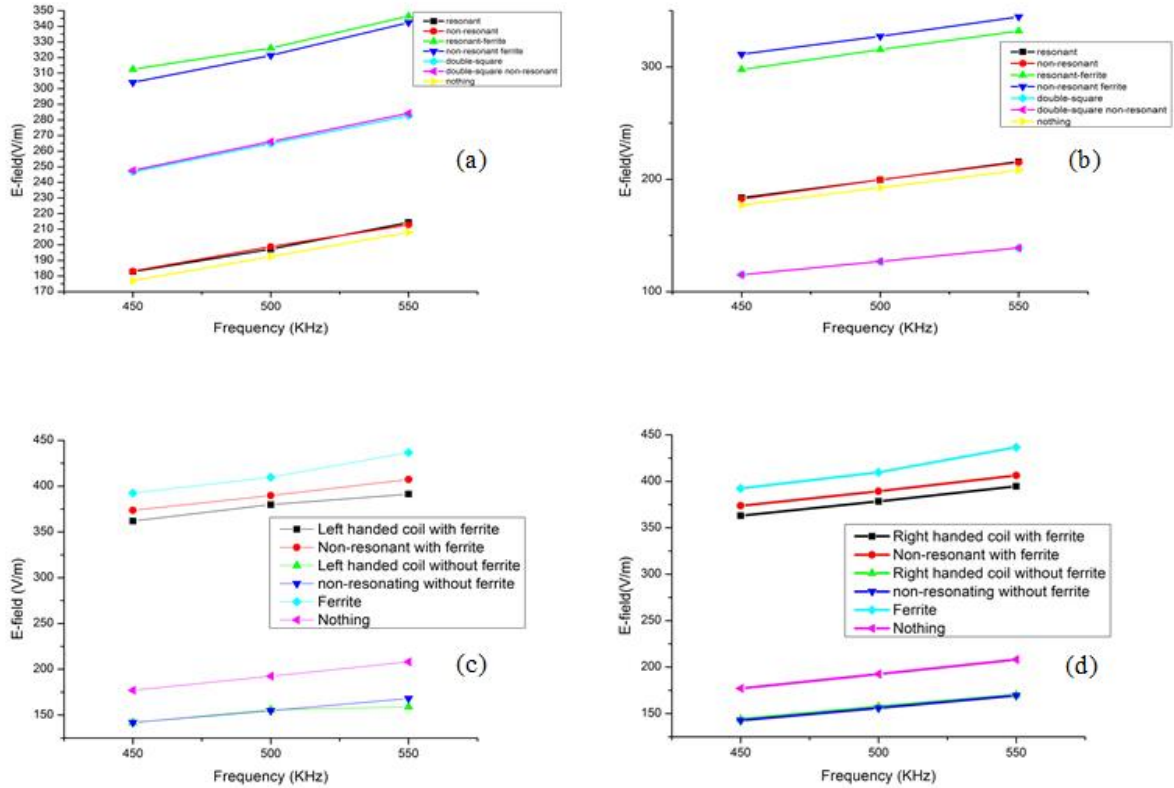

Fig. 5. Comparison studies with different dimension of coils and resonant structures (a) Left handed coils of radius 2 cm with ferrite, double square coils and vacuum placed under the tissue. (b) Right handed coils of radius 2 cm with ferrite, double square coils and vacuum placed under the tissue. (c) Left handed coils of radius 3 cm with ferrite and vacuum placed under tissue (d) Right handed coils of radius 3 cm with ferrite and vacuum placed under the tissue.

Power transfer between coils placed was found to be the maximum, when the radius of the coil was chosen to be 2 cm and hence calculations were adopted for coils and ferrite structures having radius of 2 cm. It is to be noted that the coils were wound around the ferrite core. Two terminologies are adopted in the calculation of these values. The coils which are wound in the anticlockwise direction when viewed from the source coil are called right handed coils and the coils which are wound in the clockwise direction when viewed from the source coil are called left handed coils.

The primary coil was kept constant as a double square coil. The secondary coils and cores were varied and the results for the same are indicated in Fig. 4 and Fig. 5. It is to be noted that in Fig. 5a and Fig. 5b, double square coils are used as secondary coils. For obtaining each point in Fig. 4 and Fig. 5, calculations were carried out separately in HFSS. Resonant coils were those coils which satisfied the values for the equation $f_0 = \frac{1}{2\pi\sqrt{LC}}$, where $f_0$ is the resonant frequency and L and C are the inductance and capacitance of the secondary coils with or without the ferrites. The non-resonant coils were simulated by replacing the capacitor with a short when simulation was performed.

From simulation results in Fig. 5c and Fig. 5d where ferrite cores were used, it was shown that the electric field increased by 121.71% at 450 kHz, 112.8% at 500 kHz and 109.8% at 550 kHz. Placing ferrite cores under the tissue hence, led to considerable difference in the electric field distribution. Modulation of electric field thus produced was also possible by adding resonating/ non-resonating coils in the presence or absence of these ferrite cores. Inferences obtained are stated as follows 1) power consumption can be reduced for generating the same electric field by using secondary ferrite cores, 2) Scaling of devices is possible since the power can be reduced, 3) Modulation of electric field is performed by the use of coils, 4) The theory explained earlier is validated.

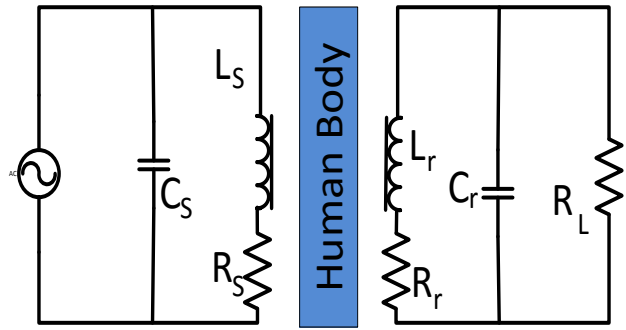

Fig. 6. Equivalent circuit diagram of the entire system

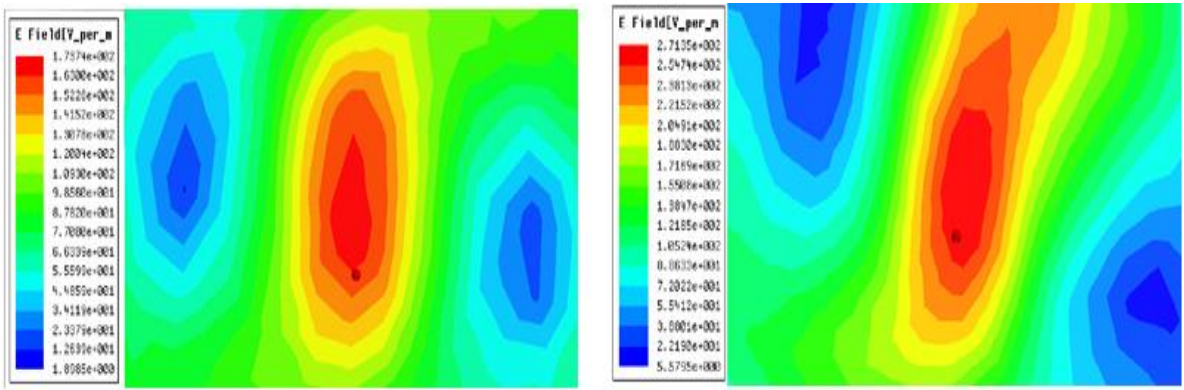

Fig. 7. Induced electric field at a frequency of 450 Khz when (Left) vacuum is placed below the tissue (b) a ferrite core of radius 2 cm is placed under the tissue

Fig/ 6. shows the equivalent circuit diagram of the system. Calculations of these individual values can be obtained by following the procedure described earlier in equations 17-19. Simulation results obtained from the electromagnetic analysis is presented in Fig. 7. for visualization of the electric fields, when vacuum and ferrite core of radius 2 cm is placed under the tissue at a frequency of 450 KHz. Fig. 8-9 depicts the measurement results using these coils which have been performed according to the set-up shown in Fig. 3. The measurements for the proposed set-up have been performed in both scenarios; by using only electrodes (with air between the electrodes) and pork tissue connected to the electrodes for measurement The use of ferrite core structures for both the above mentioned set-ups, has been found to increase the induced electric field across the electrodes by 122% as opposed to no use of ferrite cores measurement results. The results therefore, were both verified in experiment and theory for the simulation models and experimental set-ups described earlier. These are preliminary results establishing the proposed concepts. Different human body models modelled using the software and their corresponding body parts maybe used in place of the pork tissue used here, for obtaining the desired electric field to be induced. This may be potentially explored in further studies.

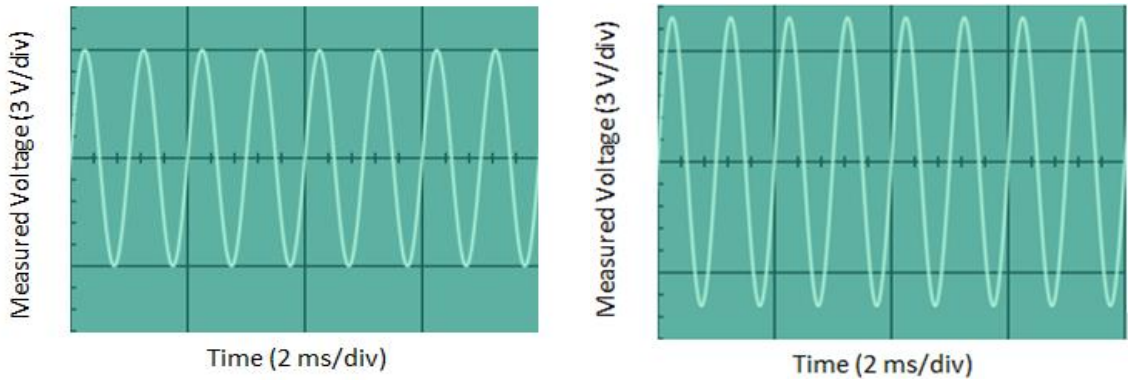

Fig. 8. Measured voltage at 450 kHz across normal electrodes (Left) without ferrite cores (Right) with ferrite cores

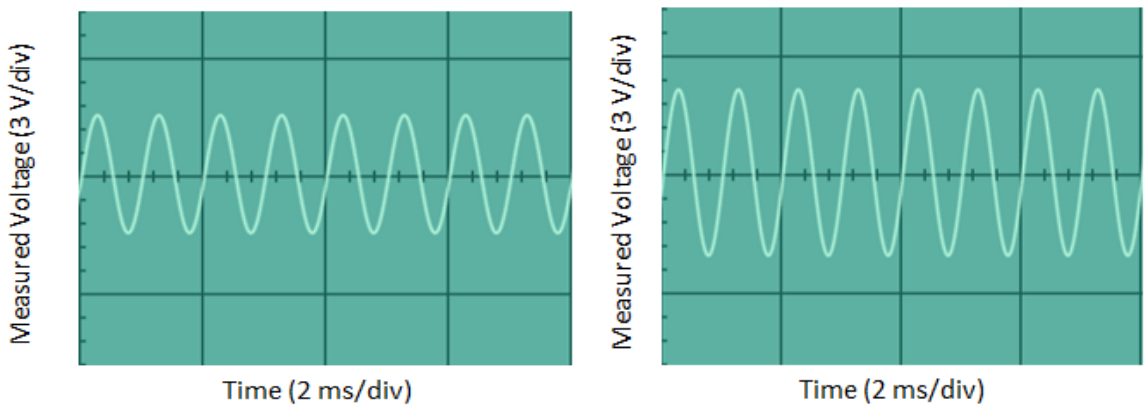

Fig. 9. Measured voltage at 450 kHz across pork tissue (Left) without ferrite cores (Right) with ferrite cores

## V.    Conclusion

In summary, mathematical modeling and analysis are carried out in the first section which aids in estimation of frequency and selection of primary coil and secondary structures. This work estimated that the electric field improved by 122% at 450 kHz when ferrite cores where placed below the tissue. It was found out that modification of the electric field is possible through the addition of secondary resonant and non-resonant coils in the absence and presence of ferrite cores. This device set-up can aid in stimulation and also can be used for carrying out experiments on the analysis of high frequency magnetic fields in the human body at reduced power consumption. The simulation has been carried out in an environment which can mimic the real scenario. Measurements are performed by using coils connected to instrumentation amplifiers are performed with normal electrodes and tissue placed in between, for both ferrite cores and no ferrite cores. The measurement results are found to be similar to that of simulation results.

Ferrite cores operating at these frequencies which are used for high power applications are available in the market. Materials like material 79 from the ferrite corporation can be used as the core. Power amplifier topologies are available in market which can provide the input power required for stimulation (AP400B by Advanced Test Equipment Corporation). The stimulator can be used for various neural stimulation applications which require modulation of electric fields in the brain, such as pain, depression, insomnia, epilepsy etc, presently being addressed using implant solutions placed within the brain . Studying of small structures like retina and the effects of magnetic fields on them can also be accomplished because of the scalability of the stimulator.

**Acknowledgments:** This research is supported by the Singapore National Research Foundation under Exploratory/ Developmental Grant (NMRC/EDG/1061/2012) and administered by the Singapore Ministry of Health's National Medical Research Council.

**Conflicts of Interest:** The authors declare no conflict of interest.

## References:

1. Esselle KP, Stuchly MA. Neural stimulation with magnetic fields: analysis of induced electric fields. *IEEE Transactions on Biomedical Engineering* **1992**; 39: 693-700.
2. Basham E, Zhi Y, Wentai L. Circuit and Coil Design for In-Vitro Magnetic Neural Stimulation Systems. *IEEE Transactions on Biomedical Circuits and Systems* **2009**; 3:321-31.
3. Rossini PM, Rosinni L, Ferreri F. Brain-Behavior Relations: Transcranial Magnetic Stimulation: A Review. *IEEE Engineering in Medicine and Biology Magazine* **2010**; 29:84-96.
4. Lefaucheur JP. Methods of therapeutic cortical stimulation. *Neurophysiologie Clinique/Clinical Neurophysiology* **2009**; 39:1-14.
5. Burunkaya M. Design and Construction of a Low Cost dsPIC Controller Based Repetitive Transcranial Magnetic Stimulator (rTMS). *J. Med. Syst.* **2010**; 34:15-24.
6. Peterchev AV, Jalinous R, Lisanby SH. A Transcranial Magnetic Stimulator Inducing Near-Rectangular Pulses With Controllable Pulse Width (cTMS). *IEEE Transactions on Biomedical Engineering* **2008**; 55: 257-66.
7. Dong-Hun K, Georghiou GE, Won C. Improved field localization in transcranial magnetic stimulation of the brain with the utilization of a conductive shield plate in the stimulator. *IEEE Transactions on Biomedical Engineering* **2006**; 53: 720-25.

8. Alkhateeb A. Gaumond RP. Excitation of frog sciatic nerve using pulsed magnetic fields effect of waveform variations. *IEEE 17th Annual Conference Proceedings in Engineering in Medicine and Biology Society* **1995**; 2: 1119-20.
9. Kotnik T, Miklavcic D. Second-order model of membrane electric field induced by alternating external electric fields. *IEEE Transactions on Biomedical Engineering* **2000**; 47: 1074-81.
10. Chen A, Moy VT. Cross-linking of cell surface receptors enhances cooperativity of molecular adhesion. *J Biophys* **2000**; 78: 2814-20.
11. Simeonova M, Gimsa J. The Influence of the Molecular Structure of Lipid Membranes on the Electric Field Distribution and Energy Absorption. *J Bioelectromagnetics* **2006**; 27: 652-66.
12. Dielectric Properties for Body Tissues. Institute for Applied Physics (IFAC) *http://niremf.ifac.cnr.it/tissprop/* 2016 [accessed 26.04.16].
13. Karalis A, Joannopoulos JD, Soljačić M. Efficient wireless non-radiative mid-range energy transfer. *Annals of Physics* **2008**; 323: 34-48.
14. Ansoft HFSS Software, http://www.ansys.com/Products/Electronics/ANSYS-HFSS.
15. Cannon BL, Hoburg JF, Stancil DD, Goldstein SC. Magnetic Resonant Coupling As a Potential Means for Wireless Power Transfer to Multiple Small Receivers. *IEEE Transactions on Power Electronics* **2009**; 24: 819-25.

## Bibliographies

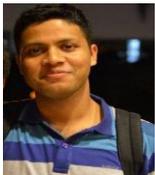

**Raunaq Pradhan** received his bachelors degree in Biomedical Engineering from National Institute of Technology, Rourkela in 2012. He has submitted his PhD thesis at Nanyang Technological University, Singapore in the year 2016 and is currently working as a research engineer at Nanyang Technological University, Singapore. His areas of interest are medical device development, magnetic/ acoustic stimulation, drug and gene delivery using magnetic fields.

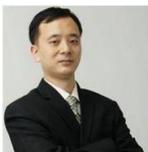

**Dr. Zheng Yuanjin** received his B.Eng. from Xi'an Jiaotong University, P. R. China in 1993, M. Eng. from Xi'an Jiaotong University, P. R. China in 1996, and Ph.D. from Nanyang Technological University, Singapore in 2001. Since July, 2009, he joined Nanyang Technological University as an assistant professor. He has been working on electromagnetic and acoustics physics and devices, biomedical imaging especially photoacoustics / thermoacoustics imaging and 3D imaging, energy harvesting circuits and systems etc.